\def\PRL#1{{\sl Phys. Rev. Lett. }{\bf #1}}
\def\PRB#1{{\sl Phys. Rev.  B }{\bf #1}}
\def\PR#1{{\sl Phys. Rev. }{\bf #1}}
\documentstyle[aps,preprint]{revtex}
\begin{document}
\draft
\preprint{}
\def\Omegahat{{\bf \hat{\Omega}}}
\def\bvec#1{{\bf \vec{#1}}}
\title{Laughlin State on Stretched and Squeezed Cylinders and\\
Edge Excitations in Quantum Hall Effect}
\author{ E. H.  Rezayi}
\address{Department of Physics and Astronomy \\
California State University at Los Angeles\\
Los Angeles, California 90032 }

\vskip 0.5in
\author{F.D.M. Haldane}
\address{
Department of Physics\\
Princeton University \\
Princeton, New Jersey 08544}

\maketitle
\newpage
\begin{abstract}
\addtolength{\baselineskip}{2ex}

We study the Laughlin wave function on the cylinder.  We find
it only describes an incompressible fluid when the two lengths
of the cylinder are comparable.  As the radius is made smaller
at fixed area, we observe a continuous transition to the charge
density wave Tao-Thouless state. We also present some exact properties
of the wave function in its polynomial form.  We then study the edge
excitations of the quantum Hall incompressible fluid modeled by the
Laughlin wave function. The exponent describing the fluctuation of
the edge predicted by recent theories is shown to be identical
with numerical calculations.  In particular, for $\nu=1/3$, we obtain
the occupation amplitudes of edge state $n(k)$ for 4-10 electron size
systems. When plotted as a function of the scaled wave vector they
become essentially free of finite-size effects.  The resulting curve
obtains a very good agreement with the appropriate infinite-size
Calogero-Sutherland model occupation numbers. Finally, we numerically
obtain $n(k)$ of the edge excitations for some pairing states which
may be relevant to the $\nu=5/2$ incompressible Hall state.
{}~~\\
{}~~PACS numbers 73.40.Kp, 73.20.Dx, 73.50.Jt \\
{}~~June 1994, v2.6\\
\end{abstract}
\newpage
\addtolength{\baselineskip}{2ex}

It is now well-established that
the quantum Hall effect (QHE) in the spin-polarized two-dimensional
electron liquid at Landau-level filling-factor $\nu$ = $1/q$
results from an incompressible correlated electron state which is very well
described by the simple
model wave functions introduced by Laughlin\cite{laughlin}.
This has been verified by extensive numerical studies\cite{qhebook}
of  systems of a finite number $N$ of electrons which
have
been carried
out in two popular geometries: spherical\cite{halehr} and periodic, or
toroidal\cite{tor1,tor}.

These geometries are convenient for obtaining bulk properties by
extrapolation to the thermodynamic limit as they do not introduce
edges.   In the most symmetric spherical geometry, the geometry is fully
specified by the requirements of translational and rotational symmetry;
in the periodic case, rotational symmetry is absent, and while the
{\it area}
of the elementary cell is fixed by the integrally-quantized
number of magnetic flux
quanta that pass through the surface, the
Bravais lattice of translations in which physical (gauge-invariant)
quantities are formulated can be continuously varied.   When the shortest of
these
translations has a length comparable to or shorter than a magnetic
length (or than the mean bulk interparticle spacing) one may expect that
the results of the finite-size calculation are no longer
representative of the bulk physics.

There are
other geometries in which the QHE has been discussed;
in particular, {\it cylindrical} geometry appears well suited for
studies of
{\it edge states} associated with the boundaries of a region of
incompressible fluid.   In this geometry, there are two characteristic
lengths in the $N$-electron system: one is the periodic repeat distance
(circumference of the ``cylinder'') which plays the role of the shortest
translation in the periodic system; the other is the width of the
finite strip of incompressible fluid that results when the electrons
are confined in the well of a potential that varies in the direction
parallel to the axis of the cylinder.   When this length becomes small,
there are interactions between the two edges of the fluid, and
calculations in this limit
(unlike the case of small periodic repeat distance)
are presumably relevant to the physics of the QHE
in narrow channels.

In this paper we describe the properties of the Laughlin
wave function
on the cylinder as the circumference of the
cylinder is
varied at fixed particle number ({\it i.e.}, fixed surface area
of the incompressible fluid) and study its' edge excitations.
As usual, we only consider electrons restricted
to states within the lowest Landau level.

The Laughlin wave function on the cylinder\cite{thouless} for $\nu=1/m$
is given by:
\begin{eqnarray}
\Psi_L
&=&
\prod_{i<j} {(e^{i\gamma z_i\over \ell}-e^{i\gamma z_j\over \ell})}^m \prod_i
e^{-y_i^2\over
2\ell^2},
\end{eqnarray}
where $\gamma = \ell/R$, $\ell = (\hbar /e B)^{\frac{1}{2}}$
is the magnetic
length, and $R$ is the radius of the cylinder.  The complex coordinate
$z$ is ${x+iy}$, where $x$ is the periodic coordinate along the
circumference of the cylinder and $y$ is along the axis.

The periodicity in the $x$-direction means that the the allowed
pseudo-momentum
$k_x$ of each particle is quantized: $k_x = n/R$, where the integer $n$
labels the conventional basis of single-particle ``Landau gauge''
states.
We allow the range of y to become
infinite, but ``compress'' the electrons by
restricting them to the Landau-gauge states with $n$ in the range
$0,1,2,\ldots,N_{\phi} $.  Just as in the spherical geometry,
the Landau level degeneracy is equal to $N_L=N_\phi+1$, which is
proportional to the area available for the guiding center motion.

It is important to note that the study of many-particle states on the
cylinder is different from other geometries because of a similar
degeneracy if the ground state is an incompressible fluid.
  In this case, the fluid will
determine its own density and need not occupy the entire surface
available to it, but will form a ``ribbon'' on the cylinder.  The
displacement of this ribbon as a whole will cause an overall
degeneracy.  For finite systems interactions with the edges will lift these
into
quasi-degeneracies.  As a result, care needs to be exercised in sorting
out the internal excitations from the translational modes.

We construct the  Laughlin wave function $\Psi_L$
numerically as the only zero energy state of the
hard-core potential with pseudo-potential parameters $V_m \ne 0$ for
$m=0,1,\ldots,\nu^{-1}-1$,  where $V_m$ are the pair energies in the
state of relative angular momentum $m$.

The Hamiltonian can be obtained by a generalization
of the pseudo-potential formulation\cite{qhebook} to the
cylinder as follows:

\begin{eqnarray}
H
&=&
{\gamma\over 2\pi }\sum_{i<j} \sum_{n,m} \int_{-\infty}^\infty
dq V_m L_m(q^2\ell^2+
\gamma^2n^2) e^{-{q^2\ell^2 \over 2}}e^{-{\gamma^2n^2\over 2}}
e^{i\gamma n \hat{y}_i\over \ell} e^{iq(\hat{x}_i-\hat{x}_j)}
e^{-i\gamma n \hat{y}_j \over
\ell}.
\end{eqnarray}
The parameter $\gamma$ serves as an ``aspect ratio'' (we fix the
area to preserve the Landau level degeneracy), and it
proves to be the crucial
parameter in determining whether the ground state is an incompressible
fluid or not. For finite systems, the physical properties of $\Psi_L$
depend strongly on $\gamma$.  For $\gamma \ge 0.7$, $\Psi_L$ becomes
different from an incompressible uniform fluid and begins to
approach a charge density wave-like state, which for larger $\gamma$
($\simeq 1.2$) is essentially described by the Tao-Thouless\cite{tt} (TT)
state.

To investigate the $\gamma$ dependence, we note that in
occupation space $\Psi_L$  has the general form:
\begin{eqnarray}
\Psi_L
&=&
\sum_{\{n_i\}}A(n_1,n_2,\ldots,n_N) \prod_i{(e^{i\gamma z_i})}^{n_i}
\prod_i e^{-y_i^2 \over 2 \ell^2} \\
&=&
\sum_{\{n_i\}}{A(n_1,n_2,\ldots,n_N)\over \{a(\gamma)\}^N}
\prod_i{(e^{\gamma^2 n_i^2 \over 2})}
\prod_i \Psi_{n_i} (z_i)
\end{eqnarray}.
$A(n_1,n_2,\ldots,n_N)$ are amplitudes {\em independent} of $\gamma$,
$a(\gamma)$ is an overall normalization factor {\em independent}
of the occupation numbers, and $ \Psi_n(z)$ are the appropriate single
particle wave functions. An algorithm for calculating these
amplitudes for small system sizes has been given previously\cite{ishi}, but
this does not appear to
provide any substantial advantage to the alternative of diagonalizing
H directly.  Accordingly, we calculate $\Psi_L$ numerically from $H$ for
$\gamma=0$ and then construct it for $\gamma\ne 0$ from Eq. 4. The
charge densities shown in Fig. 1 and 2 were obtained in this manner.  The
parameter $a$ in the figures is the aspect ratio:
$a=L/(2\pi R)=(\gamma^2\times N_L)/2\pi$,
where $L$ is the length of the cylinder.
  The square points in Figs. 1 and 2 are the
average occupation of each orbital positioned at the center of the
Gaussian envelope of the orbital,
which are separated by $\Delta y=\gamma$.  These figures capture
the transition to the charge density wave state
as $\gamma$ is increased.

The extremely squeezed cylinder (hoop geometry) is realized at $\gamma=0$; this
is an interesting limit where the spatial dimensions have been reduced to
essentially one in that the Gaussian envelopes  are always centered at $y=0$.
The density however does have a finite width in the squeezed direction (see
Fig. 3).
 We have found several interesting properties of the Laughlin's wave function
for $\gamma =0$ which we now address. In this
limit,  $\Psi_L$  assumes the familiar form:
\begin{eqnarray}
\Psi_L^{\gamma=0}
&=& \prod_{i<j} {(z_i-z_j)}^m \prod_i e^{-y_i^2 \over 2 \ell^2}.
\end{eqnarray}
Clearly in the expansion of this wave function all occupation
amplitudes are integers.  In particular, expanding the Jastrow factor we
obtain:
\begin{eqnarray}
\prod_{i<j} (z_i-z_j)^m
& = & \sum_{m_1,m_2,\ldots,m_n} C(m_1,m_2,\ldots,m_N) \prod_{i=1}^N
z_i^{m_i}.
\end{eqnarray}
In the appendix, we show that:
\begin{itemize}
\item{(a)} $C((\{m_{Pi}\})=((-1)^P)^m C((\{m_i\})$, $P$ is a permutation
on $N$ objects.

\item{(b)} $C(\{m_i^0\})=1$, where $\{m_i^0\}=\{m(i-1)\}$.

\item{(c)} $C(\{m_i\}) = m \times$ integer, (integer can be zero) if $\{m_i\}$
can be
obtained from a permutation of $\{m_i^0\}$ by a succession of
``squeezing'' operations (defined below).

\item{(d)} $C(\{m_i\}) =0$ if $\{m_i\}$ is not a permutation of
$\{m_i^0\}$ or a configuration obtained by successive ``squeezing''
of $\{m_i^0\}$.
\end{itemize}

The squeezing operation (without  reordering) $\{m_i\} \rightarrow
\{m_i^\prime\}$ is defined
for some pair $ j \ne k$ with $m_j > m_k$ and a positive integer $n$ by:

$
\begin{array}{cccc}
m^\prime_i \ \ =\ \  m_i&\mbox{if, $i \ne j,k$}&&\\
\left.
\begin{array}{ccc}
m_j^\prime &=& m_{j}-n\\
m_k^\prime&=&m_{k}+n\end{array}
\right \}& \mbox{where $0<n<m_j-m_k$.}&&
\end{array}
$

Note that: $$\sum_{i < j} (m_i-m_j)^2 > \sum_{i <j } (m_i ^\prime-m_j
^\prime )^2 = \sum_{i < j} (m_i-m_j)^2 + 2nN(n-(m_j-m_k)).$$
Since every occupation state with non-zero amplitude
can be squeezed from the TT state and, since within constants, $\sum_i n_i^2$
is related to $\sum_{i<j} (n_i-n_j)^2$, it can be seen from Eq. 4 that the
above  inequality implies
that the TT state will dominate the wave function as
$\gamma \rightarrow \infty$.
In practice this
condition is realized for $\gamma\ge 1$, see Fig. 2.

We have also empirically verified the following for up to $N=8$
electron systems, the proof of which we leave as an open problem:
\begin{itemize}
\item{(1)} The filled level droplet has the highest amplitude.
\item{(2)} $A(\mbox{filled level droplet})={ (-1)^{[{N+2 \over m-1}]}[{m+1\over
2} N]! \over N![ {m+1 \over 2} ]!^N}$

\item{(3)} The total normalization is $||\Psi||^2 = {[mN]! \over N!
{[m!]}^N}$.
\end{itemize}

The first four items (a-d) are also valid for bosons except of course
the wave function
is now
symmetric.  The last three items (1-3) are however only valid for fermions.

It should be noted here that in the squeezed limit the appropriate 1-d
Hamiltonian for which the $\nu=1/3$ Laughlin state is an exact zero-energy
ground state is:
\begin{eqnarray}
H(x_1,x_2,\cdots,x_N)= - \sum_{i<j} {\partial^2 \over \partial x_i^2 } \delta
(x_i-x_j)
\end{eqnarray}
This result also noticed by Wen is an
obvious limit of the full two-dimensional hard core
potential\cite{qhebook,Trug}.

Before we study edge states we would like to make a few comments.  The period
of the oscillation
of the density near the edges in the
fluid phase is determined by the bulk density-density response function which
is
dominated by a single mode namely the magneto-roton\cite{GMP}.
As such, it is similar to the impurity screening (or the lack of it) by
the incompressible fluid.  The period at the edge
is comparable to that of
the density oscillations near an impurity\cite{impur}. In the limit of large
aspect ratios, the system makes a transition to a ``stripped Wigner Crystal''
phase where the number of peaks in the density is equal to number of electrons.
This same limit, but for the Coulomb interactions,
has been studied by Chui\cite{chu} in connection with narrow
channels. However, he used toroidal boundary conditions.  It is not surprising
to see gapless modes\cite{foot} (as we see) in this limit.
In the opposite limit,
e.g. small $\gamma$, the gap survives at least for the hard-core
repulsion and may be the more appropriate limit for narrow channel geometries.
Whether or not QHE can be ruled out for very narrow channels has not
yet been addressed in our study and needs further investigation.

We next turn to the discussion of edge
excitations of the Laughlin droplet (or ribbon in our case).  Again,
the most interesting limit turns out to be the completely squeezed
cylinder ($\gamma=0$).  In this limit there is no bulk in the usual
sense and only edge excitations remain.
Wen\cite{wen} has used current algebra techniques to
study edge states with the same results as here.  One of
us\cite{duncvarena} has interpreted the edge states as
``generalized Fermi surface singularities'', obeying a generalized
Luttinger\cite{Lutt} theorem.  Here we follow that formalism.

Consider first a free Fermi gas with a momentum space energy surface
$E(k)$ having the topology shown in Fig. 4.  In terms of the Fermi points
$k_f^i$, we have the Luttinger theorem for the total number of particles
$N$ and the total momentum of the system $P$:
\begin{eqnarray}
N&=&{L\over 2\pi} \sum_i\Delta \nu_i k_f^i\\
P&=&{L\over 4\pi} \sum_i \Delta \nu_i (k_f^i)^2
\end{eqnarray}
where $\Delta \nu = n(k_f-\epsilon)-n(k_f+\epsilon)$ is the jump across
the Fermi point, and $L$ is the length of the system.  One can
generalize
these expressions to contain local deviations of the Fermi points
$\delta k_f^i(x)$ from their uniform value $k_f^i$:
\begin{eqnarray}
\Delta \rho(x)& =&{1\over 2\pi} \sum_i\Delta \nu_i \delta k_f^i(x)\\
\Delta \Pi (x)&=&{1\over 2\pi} \sum_i \Delta \nu_i k_f^i \delta k_f^i(x)+
{1\over 4\pi} \sum_i \Delta \nu_i (\delta k_f^i(x))^2
\end{eqnarray}
where,
\begin{eqnarray}
k_f^i(x)&=&k_f^i+{2\pi\rho_i(x)\over \Delta \nu_i}\\
\rho(x)&=&N\over L \\
\Pi(x)&=&{P\over L}.
\end{eqnarray}
The local description is valid only if for any pairs of indices $i,j$ the
condition
$(k_f^i-k_f^j)\xi \gg 1$ is satisfied. Here $\xi$ is an inverse
momentum cut-off measured from the
Fermi points.

Similarly the Hamiltonian for edge deformation
can be expressed in terms of the local fluctuations
of the Fermi wave vector:
\begin{eqnarray}
H^{(0)}&=&{1\over 4\pi} \int dx \sum_i v_f^i \Delta \nu_i (\delta k_f^i(x))^2
\end{eqnarray}
where $v_f^i$'s are the Fermi velocities.  It is possible to include
interactions terms between these fluctuations as in Fermi Liquid theory
but such terms are not relevant to this work and will not be pursued here.
We next proceed to quantization.

It is well known that for 1-d Fermi gas\cite{Solyom},
the Fourier components of the density operators (Tomonaga bosons)
\begin{eqnarray}
\rho^i(x)&=&{1\over L} \sum_{|q|<\xi^{-1}} e^{iqx} \rho_q^i
\end{eqnarray}
satisfy the Kac-Moody algebra:
\begin{eqnarray}
[\rho_q^i,\rho_{q\prime}^j]& =& \Delta \nu_i \delta_{i,j} {qL \over 2\pi}
\delta_{q+q\prime,0},\ \ \ \ \ \ q\xi \ll 1.
\end{eqnarray}
Furthermore, Bosonizations of Fermi fields in 1-d give:
\begin{eqnarray}
\Psi_i^\dagger (x)&=& A_i e^{i\int_{-\infty}^x dx^\prime k_f^i(x^\prime)} \\
                & =& A_i e^{i\phi_i(x)}
\end{eqnarray}
where $A_i$, to be given shortly, are operators which make $\Psi$'s
anticommute,
and $\phi_i(x)$ is the canonical Bose field.  From these and Eq. 12 we obtain:
\begin{eqnarray}
{\partial \phi_i \over \partial x}&=&k_f^i(x) \\
&=&k_f^i+{2\pi\rho_i(x)\over
\Delta \nu_i}
\end{eqnarray}
which is the ``chiral constraint''.  Next, from the above Eqs., we obtain the
commutation rules (CR):
\begin{eqnarray}
[\rho_i(x),\rho_j(x^\prime)]&=&\delta_{i,j} {\Delta \nu_i \over 2\pi}{d\over
dx}
(\delta(x-x^\prime))
\end{eqnarray}
\begin{eqnarray}
[\rho_i(x),{\partial \phi (x^\prime)\over \partial x^\prime}]&=&\delta_{i,j}
{d\over dx} (\delta(x-x^\prime))
\end{eqnarray}
\begin{eqnarray}
[\phi_i(x),\rho_j(x^\prime)]&=&i \delta_{i,j} \delta(x-x^\prime)
\end{eqnarray}
\begin{eqnarray}
[\phi_i(x),{\partial \phi(x^\prime)\over \partial x^\prime}]&=&
2\pi i \Delta \nu_i \delta_{i,j} \delta(x-x^\prime)
\end{eqnarray}
\begin{eqnarray}
[\phi_i(x),\phi_j(x^\prime)]&=& {i\pi \over \Delta \nu_i} \delta_{i,j}
Sgn(x-x^\prime)
\end{eqnarray}
Thus $\phi$, and $\rho$ (or the derivative of $\phi$) are conjugate
fields.  Using These CR's and
\begin{eqnarray}
[N,\Psi_i^\dagger(x)]&=&\Psi_i^\dagger(x),
\end{eqnarray}
it can be seen that:
\begin{eqnarray}
\Psi_i^\dagger(x)\Psi_i(x^\prime)&=&\Psi_i(x^\prime)\Psi_i^\dagger(x)
e^{-[\phi_i(x),\phi_i(x^\prime)]}.
\end{eqnarray}
For the fields to obey Fermi statistics we must require that
${1\over \Delta \nu}$ be an odd integer $\pm m$.  We also note that
the choice
\begin{eqnarray}
A_i& =& e^{{i\pi\over 2} \sum_j Sgn(i-j)N_j}
\end{eqnarray}
makes the Fermi fields for different sectors ($i\neq j$) anticommute as well.

The usual results for the 1-d Fermi gas are recovered when
$\Delta \nu =1$. On the other hand,  applying this formalism to
the quantum Hall effect we obtain
$\Delta \nu =1/m$.  The CR's for the Bose fields $\phi(x)$ in these two
cases are
therefore related by:
\begin{eqnarray}
[\phi(x),\phi(x^\prime)]_{\mbox{QHE}}& =&{1\over m} [\phi(x),\phi(x^\prime)]_{
\mbox{Fermi}}
\end{eqnarray}
Following standard treatment of the Luttinger model\cite{Solyom}
 ,except here one  needs to consider only a single branch of fermions
 (for example the right moving one),
the single particle Green's function is:
\begin{eqnarray}
G(x-x^\prime,t-t^\prime)& =& ({1\over x-x^\prime-v_f(t-t^\prime)})^m
e^{imk_f(x-x^\prime)}
\end{eqnarray}
Fourier transforming the equal time Green's function we obtain the
average occupations:
\begin{eqnarray}
n(k)& \approx & C {(k-mk_f)^m\over |k-mk_f|} + n^{\mbox{reg}}(k),
\end{eqnarray}
where $C$ is a numerical coefficients and $n^{\mbox{reg}}(k)$ is
the non-singular part of $n(k)$.  The average occupation of the
edge excitations must therefore exhibit a power law singularity
at $mk_f$ with exponent $m-1$.  We next present numerical results.
{}From here on we will only consider the completely squeezed geometry
$\gamma=0$.

Fig. 5 shows n(k) for up to 10 electron size systems
for $\nu=1/3$.  As can be seen, there is a remarkable degree of convergence
and the shape of $n(k)$ is already clear.
We have defined $k_f(N)$ for each size by having exactly $N$-states
between $-k_f \le k \le k_f$.  For each size we have rescaled k in
Fig. 5 by $k_f$.  The solid  curve
is $n(k)$ of the Calogero-Sutherland\cite{cal}
model.
It was obtained from
a conjecture by Haldane\cite{hal0} giving the full analytical
expression
of the retarded Green's function for the Calogero-Sutherland
model at integer coupling ($\lambda=m$ for $\nu=1/m$).
Haldane's expression reduces to the equal time Green's function
recently calculated by
Forrester.\cite{for}
Note that in Fig. 5 there appears to be a weakly singular
pseudo-Fermi surface at $k_f$.
 The singularity of interest however is not at
$k_f$ but
at $mk_f$, i.e. $3k_f$ in this case.  Unfortunately, even $N=10$
is not sufficiently large to make a direct extrapolation of the
exponent possible.  That is, we have not approached the $k=3k_f$
sufficiently closely (see Fig. 5).  However, we
 extract the exponent as follows: we observe that
the last possible $k$ for which $n(k)$ is non-zero satisfies
$(k_{max}-3k_f) \sim 1 /N$ (see below).  Now if $n(k) \sim (k-3k_f)^2$
then $n(k_{max}) \sim  1/N^2$, which implies that,
in the thermodynamic limit the quantity
$N^2n(k_{max})$ should extrapolate to a finite value.  Fig. 6
shows such an extrapolations clearly demonstrating the $(k-3k_f)^2$
dependence.  The value of exponent has also been confirmed in the planar
disk
geometry\cite{mac}.

At first sight, the data in Fig. 5 seems to be size independent and to
exactly form a universal curve.  This is not so, there are extremely
small finite-size effects.
In fact, we have empirically obtained the exact expression for
the last 4 of $n(k)$ in the tail
of the distribution
as a function of system size:
\begin{eqnarray*}
n_0&=& {2 \over 2+9N(N-1)} \\
n_{-1}&=& {18(N-1)\over (2+9N(N-1))(1+3(N-2))} \\
n_{-2}&=& {36(N-1)(9N^2-30N+22)\over
(2+9N(N-1))(2+9(N-1)(N-2))(1+3(N-3))} \\
n_{-3}&=&{4(2+3(N-3))(405N^4-2970N^3+7425N^2-7554N+2698)\over
(2+9N(N-1))(2+9(N-1)(N-2))(2+9(N-2)(N-3))(1+3(N-4))}
\end{eqnarray*}
where $n_0\equiv n(k_{max})$, $n_{-1}\equiv n(k_{max-1})$, etc.
One might wonder if there is a pattern to be used for constructing
successively higher terms.
While the factors in the denominator appear to follow a simple
pattern,  it is clear that
the one polynomial in the numerator rapidly becomes complicated
and is not easily generalized.  However
it should be noted that, as $N\rightarrow\infty$, the ratio of these
coefficients become very simple:
\begin{eqnarray}
{n_{0}\over n_{0}}& =& 1 \\
{n_{-1}\over n_{0}}& =& 3 \\
{n_{-2}\over n_{0}}&=&6 \\
{n_{-3}\over n_{0}}&=&10.
\end{eqnarray}
This sequence makes it clear that these ratios are just
the expansion coefficient of $(1-x)^{-3}$,  That is:
$${n_{-p} \over n_0} ={(p+1)(p+2)\over 2}$$
This particular increasing sequence has also been
noticed by Wen\cite{wenp}.  We note in
passing that
the simple extrapolation procedure in Fig. 6, where we essentially extend
the line connecting the points $N=10$ and $N=9$, out to $N=\infty$,
gives the value of 0.22 for the coefficient of $N^{-2}$, which is
remarkably close to the exact 2/9 answer.  It should also be noted here that
one
can directly extract the exponent from the analytical expression of
$n_0$ by using the identity
$$ {k-mk_f\over k_f} = -{m\over N}\ \ \ \ (m=3)$$ to eliminate $N$ in favor of
$k-3k_f$.

We have also calculated the average occupations for the
Boson Laughlin state at $\nu=1/2$. Here the the exact analytical form of
$n(k)$ is known\cite{dyson}
\begin{eqnarray}
n(k)&=&C^\prime \Theta(2k_f-k)\ln {2k_f\over k},
\end{eqnarray}
where $C^\prime$ is another constant and $\Theta$ is the step function.  Note
that the logarithmic dependence upon $k$
predicts a linear relation between $n(k)$
and $k-2k_f$, which is clearly seen in Fig.  7.
It should also be noted that the increase of
$n(k)$ with size at $k=0$ is highly suggestive of a weak singularity,
consistent with
the logarithmic form in Eq. 33.
This singularity is however an artifact of the squeezed
limit and will disappear for non-zero $\gamma$ as will the singularity
at $k_f$ in the Fermi case.

There are, in the realm of the quantum Hall effect, other
states with perhaps  more interesting edge excitations.  We will
not address all such states here but confine our attention to the
pairing ones which have been proposed as candidates for the
$\nu=5/2$ Hall state.  First a spin-singlet
state was proposed by us\cite{HR}:
\begin{eqnarray}
\Psi_{HR}&=&\prod_{i<j}(z_i^\uparrow-z_j^\uparrow)^2 \prod_{i<j}(z_i^\downarrow
-z_j^\downarrow)^2 \prod_{i,j} (z_i^\uparrow-z_j^\downarrow)^2 {\bf Det} \{
{1\over (z_i^\uparrow-z_j^\downarrow)^2} \},
\end{eqnarray}
where Det stands for the determinant.

The other a spin-polarized state was proposed by Moore and Read\cite{MR} and
later studied by Greiter, Wen and Wilczek\cite{GWW}.

\begin{eqnarray}
\Psi_{MR} &=&\prod_{i<j} (z_i-z_j)^2
{\bf Pf} \{  {1\over
(z_i-z_j)} \},
\end{eqnarray}

here Pf
denotes the Pfaffian. Both the determinant and the Pfaffian act to create pairs
and are closely related to BCS wave function
for singlet and triplet pairing respectively.  One therefore might
expect as in BCS theory a smoothly falling $n(k)$ with no semblance of
any singularity at $k_f$ that was seen for the 1/3 state.  This is
clearly borne
out in
 Fig. 8 showing $n(k)$ for the spin-singlet state for
up to 8 electrons.  There is indeed a marked difference from that
of the Laughlin wave function. There is evidently, as before,
a reasonable degree of convergence
already with these small sizes.

Analogous results for the spin-polarized Pfaffian state\cite{MR,GWW}
is shown in
Fig. 9.
There is a rather sharp drop at $k_f$
possibly indicating a weak singularity, i.e. a pseudo-Fermi surface, which
is probably an artifact of the squeezed limit.
In addition, it shows a very slight dip at $k=0$.
Despite their  common pairing nature,
the two $n(k)$ of Fig. 8 and 9 appear to be different.  Presumably
because of the differences in even (spin-singlet) or odd pairing states of
the two.
A  more detailed
understanding of these differences will probably have to await until
a more precise analytical expression
for $n(k)$ is obtained.

In closing we emphasize again that
while for convenience we have studied the edge states
in the squeezed limit $\gamma=0$, our results on the edge singularity
remain valid for all
aspect ratios $\gamma \ne 0$  so long as the bulk is an incompressible fluid.
As stated earlier only the
apparent singularities
at $k=k_f$ for fermions and $k=0$ for bosons are washed out
at non-zero $\gamma$.

This work was supported by NSF Grants  DMR-9113876 at CSULA,
and DMR-8901985
at Princeton University.
\eject

{\large \bf Appendix A}

The first property follows from the antisymmetry of the wave function.
We find the state with the smallest amplitude
by choosing the configuration with lowest power of $z_1$, then lowest
power of $z_2$, etc.  We obtain:
\begin{eqnarray}
(-)^N\prod_{j=2}^N z_j^m \prod_{j=3}^N z_j^m \ldots z_N^{m(N-1)}
&=& (-)^N\prod_{j=2}^N z_j^{m(j-1)}
\end{eqnarray}

{}From this we immediately pick out the occupation numbers:
\begin{eqnarray}
n_j
&=& 1\ \ \ \ \ \ \mbox{for}\  j=0,m,2m,\ldots, Nm.\\
n_j
&=& 0\ \ \ \ \ \ \mbox{otherwise}.
\end{eqnarray}
Thus, every m$^{\mbox{th}}$ orbital is occupied, which is the TT state.
Note that TT state in configuration space can be obtained by
antisymmetrization of this expression (see below).
There is another more direct way of seeing this which also will prove
point c. We expand the polynomial in the following manner:
\begin{eqnarray}
\prod_{i<j} (z_i-z_j)^m
&=& \prod_{i<j} (z_i^m-z_j^m +mX_{i,j})\\
& & \prod_{i<j}(z_i^m-z_j^m)+m\prod_{i<j}G_{i,j}
\end{eqnarray}
The first term is the TT state with a coefficient of unity. $X_{i,j}$
and $G_{i,j}$ are appropriate remainder terms.  It can be seen from the
binomial expansion that a factor of $m$ multiplies $X_{i,j}$ and hence
$\prod_{i<j} G_{i,j}$.  This proves point c.

To establish the ``squeezability'' we write:
\begin{eqnarray}
\prod_{i<j} (z_i-z_j)^m
&=&\prod_j^{N-1} z_j^{m(N-j)}\prod_{i<j}(1-{z_j \over z_i})^m
\end{eqnarray}

Note the prefactor is related to a similar expression given above (Eq.
7) by the
permutation $\left ( \begin{array}{llll}1&2&\ldots&N\\N&N-1&\dots&
1\end{array} \right)$, with the correct sign $(-)^N$, and thus it describes
the TT state. Now consider any
pair $k,l$ with $k>l$, since in the prefactor $z_l$ is raised to a
higher power than $z_k$, but in the product $\prod_{i<j} (1-z_i/z_j)$,
$j$ is greater than $i$, it then follows that the power of $z_l$ will
increase while
the
power of $z_k$ will decrease, indicating the pair is being squeezed
together.

\vfill\eject

\newpage
\begin{figure}
\caption{The charge density plotted versus $y$ the dimension
along the length of the cylinder for $N=10$.  Also plotted are the average
occupations of the Landau level orbital placed at the position of
the guiding centers.  The aspect ratios $a={L \over 2\pi R}$ are also
indicated, in parenthesis,
on each plot.  The
series of plots show the evolution of the incompressible fluid
from the squeezed cylinder in the lower left corner as $\gamma$ is
increased counter-clockwise.}
\end{figure}

\begin{figure}
\caption{Same as figure 1 except $\gamma$ is increased further,
stretching the cylinder. The charge-density wave state starts to
appear at an aspect ratio
of 2 and is fully developed at 5.}
\end{figure}

\begin{figure}
\caption{The charge density of the completely squeezed cylinder.
The occupation amplitudes are now centered at $y=0$ since no motion
of the Gaussian envelope is possible in the squeezed direction. The
units of the density in this figure are arbitrary and independent
of those in Figs. 1 and 2. $N=10$ in this case as well.}
\end{figure}

\begin{figure}
\caption{An arbitrary energy versus momentum dispersion relation
 $E(k)$ for a 1-d
Fermi gas. The Fermi energy is shown by the horizontal line and
the Fermi points are labeled.}
\end{figure}

\begin{figure}
\caption{The average occupation of $n(k)$ versus $k$ for the
Laughlin state at 1/3 filling for up to 10 electron size systems.  The
$k$-axis has been normalized by $k_f$ so that $k/k_f=2 m_i/N$, $\ \
m_i=-m(N-1)/2,\cdots,m(N-1)/2$.  The solid curve is the exact
$n(k)$ for the Calogero-Sutherland model.}
\end{figure}

\begin{figure}
\caption{The plot of $N^2n(k_{max})$ versus $1/N$.
An
estimate of $N^2n(k_{max})$ as $N\rightarrow \infty$, based
on the extrapolation of the data,
yields 0.22.}
\end{figure}

\begin{figure}
\caption{Same as Fig. 5 but for bosons at $\nu=1/2$.  Here
there is a hint of the logarithmic singularity in the exact result
for $N\rightarrow \infty$ (see text) at $k=0$.  The logarithmic
dependence yields a linear exponent at $2k_f$ which can be seen in
this plot.}
\end{figure}

\begin{figure}
\caption{Same as Fig. 5 except for the spin-singlet pairing
state of Eq. 34.  $n(k)$ is similar to the single particle
occupation in the BCS ground state with no singularity at $k_f$.}
\end{figure}

\begin{figure}
\caption{Same as Fig. 8, but for the Pfaffian pairing state.  The diffrence is
presumably due to S-wave(Fig. 8) versus P-wave pairing.}
\end{figure}
\end{document}